\documentclass[aps,preprint,pra,showpacs]{revtex4}
\usepackage{graphicx}
\usepackage[pdftex]{color}
\def\xt{({\bf x},t)}

\def\A{{\bf A}}

\begin{document}

\title{Comment on ``Role of Potentials in the Aharonov-Bohm Effect''}

\author{Yakir Aharonov}
\affiliation{School of Physics and Astronomy, Tel Aviv University, Tel Aviv 6997801, Israel\\ and Schmid College of Science, Chapman University, Orange, CA 92866}

\author{Eliahu Cohen}
\affiliation{School of Physics and Astronomy, Tel Aviv University, Tel Aviv 6997801, Israel}

\author{Daniel Rohrlich}
\affiliation{Department of Physics, Ben Gurion University of the Negev, Beersheba
8410501 Israel}

\date{\today}

\begin{abstract}
Are the electromagnetic scalar and vector potentials dispensable? Lev Vaidman \cite{lev} has suggested that local interactions of gauge-invariant quantities, e.g. magnetic torques, suffice for the description of all quantum electromagnetic phenomena. We analyze six thought experiments that challenge this suggestion.  All of them have explanations in terms of $local$ interactions of gauge-$dependent$ quantities, in addition, some have explanations in terms of $nonlocal$ interactions of gauge-$invariant$ quantities. We claim, however, that two of our examples have no gauge-invariant formal description and that, in general, no $local$ description can dispense with electromagnetic potentials.
\end{abstract}


\maketitle


The conventional statement of the Aharonov-Bohm \cite{ab} (AB) effect is that, while electromagnetic scalar $V\xt$ and vector ${\bf A}\xt$ potentials are mere calculational aids in classical mechanics, in quantum mechanics they are an essential part of the formalism:  a charged quantum particle can interact with, and respond to, electromagnetic potentials, without ever passing through an electromagnetic field.  At the same time, only gauge-invariant quantities are measurable, and quantum mechanics is manifestly gauge invariant.  It is, therefore, natural to try to dispense with electromagnetic potentials; yet attempts to do so, over the years, have been unsuccessful.


Recently, Vaidman \cite{lev} has proposed an explanation (related to \cite{fr}) for the AB effect, via forces rather than via electromagnetic potentials. For the magnetic effect, he considers a solenoid made of two counter-rotating, oppositely charged coaxial cylinders.  He notes that even if the magnetic field of the solenoid is screened from the electron diffracting around it, the transient magnetic field of the passing electron, which is not screened from the rotating cylinders, either increases or decreases the relative rotation rate of the cylinders, according to whether the electron passed on one side or the other of the solenoid.  The overall wave function of the electron and solenoid is a superposition of two terms, one for each electron path (with corresponding solenoid motion), and their relative phase---the AB phase---is proportional to the torques induced by the transient magnetic fields integrated over the angular displacements of the cylinders.  Are the potentials, then, dispensable in this case?  Our claim in this Comment is that as part of the formalism of quantum mechanics they are {\it not} dispensable; and that as an explanation of the dynamics they may be dispensable, but only if locality, as well, is dispensable.  Below, we present six examples that make Vaidman's proposal implausible.

As a prelude to our examples, let us consider replacing the solenoid with a spatially constant magnetic field that persists briefly after its source has vanished.  According to an analysis based on local forces, the AB phase arises from the local action of the electron on the source of magnetic flux.  So if an electron interferes with itself in a magnetic field that no longer has a source, it should not acquire an AB phase.  According to quantum mechanics, the electron will indeed acquire an AB phase.  One could try to explain the AB phase as arising from interaction of the electron with photons making up the magnetic field, but the photons would have low frequencies and momenta, hence large position uncertainties, and presumably would not yield phase effects. Moreover, a detailed analysis within QED would require an interaction between the electron's fermionic field and the gauge-dependent electromagnetic potential of the photons.

Now our first three examples do not include any force at all upon which to base an explanation of the AB effect.  First, we consider a topological model of the effect.  Let a system of two particles in two dimensions have the following Lagrangian:
\begin{equation}
L={1\over 2}{m_1v_1^2}+ {1\over 2}{m_2v_2^2}+({\bf v_1}-{\bf v_2})\cdot
\A({\bf r_1}-{\bf r_2})~~~~.
\end{equation}
To define a topological effect with no forces at all, we take $\A=\nabla (\alpha \phi_{12})$, where $\alpha$ is a constant and $\phi_{12}$ is the angle between ${\bf r}_2 -{\bf r}_1$ and the positive $x$-axis. It is clear from our choice that both ${\bf E}$ and ${\bf B}$ vanish (since, locally, ${\bf A}$ is a pure gauge) except at the positions ${\bf r}_1$ and ${\bf r}_2$, where they are unphysical, since they diverge.  Yet, when one particle encircles the other, the system acquires a phase $2\pi\alpha$. The fact that there is no source here that could be subject to a local (re)action raises doubts about an explanation via local forces such as the one in \cite{lev}.  The electromagnetic potential seems essential for describing the dynamics.

\begin{figure}
\centerline{
\includegraphics*[width=120mm]{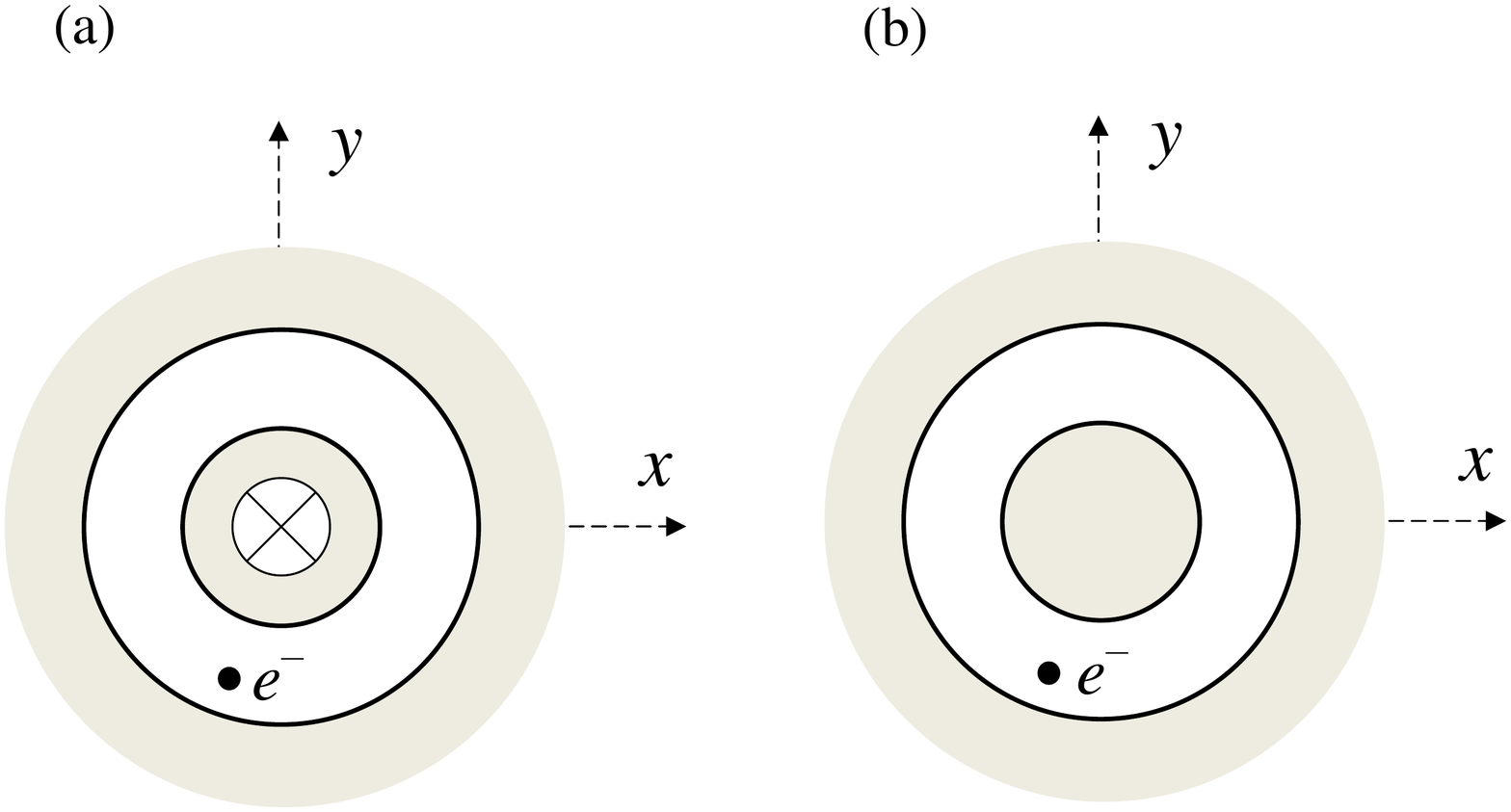}}
\caption[]{An electron confined to an annular region.  A magnetic flux (a) may or (b) may not thread the inside region, centered on the origin, where the electron cannot go.}
\label{Fig1ab}
\end{figure}

The second example involves a charged particle, say an electron, restricted to the inside of a toroid or, for simplicity, a two-dimensional annular region.  (See Fig. \ref{Fig1ab}.)  The potential $V(x,y)$ (defining the walls of the annular region) is symmetric under rotation about the $z$-axis.  Aside from $V(x,y)$ the electron is free, not subject to any force, and in its ground state.  What, indeed, is its ground state?  By virtue of the rotational symmetry, the angular momentum $L_z$ should be a good quantum number $n\hbar$, with $n$ an integer and $n=0$ in the ground state.  Note that we have not specified whether any magnetic flux threads the annular region.  Why should we?  A flux threading the annular region, as long as the electron never encounters it, cannot exert a force on the electron. And since the electron is in its ground state, it should not exert any force on the source of the flux.  But here, in particular, quantum mechanics ignores explanations based on forces.  If there is magnetic flux $\Phi_B$ threading the annulus and it is not a multiple of the flux quantum $hc/e$, then the electron has nonzero angular velocity and its energy depends on $\Phi_{B}$.  How can we explain this effect using forces, when there are no forces?

We could answer this question as follows.  We adopt the rule that, to find the ground state of a system in the presence of electromagnetic fields, we first set all the fields everywhere to zero.  We thus obtain the ground state of the system in the absence of electromagnetic fields.  To obtain the ground state in the presence of the fields, we turn them on adiabatically so that they modify the ground state.  We thus obtain the ground state in the presence of fields.  For the example of the electron in the annulus, this rule tells us that its ground state in the absence of flux has $L_z =0$; and that, when the flux changes to some value different from 0 and different from $nhc/e$, the electron's angular velocity changes to ${\dot \theta} \ne 0$, as the changing flux induces a circulating electric field in the annulus, which in turn accelerates the electron around the origin \cite{levpc}.

We thus describe the ground state of this electron without reference to potentials.  However, the price of this description is a questionable rule singling out the field-free description as the correct starting point.  What is the physical meaning of this rule?

Our third example \cite{ac} resembles the second in its topology.  This time there is no flux; indeed, there is no electromagnetic field at all, other than the field of an electron in a wave function $\psi_n (r,\theta)$ that is an eigenstate of $L_z = -i\hbar \partial /\partial \theta$ with eigenvalue $n\hbar$ for integer $n$.  In Fig. 2(a), the quantization of angular momentum component $L_z$ derives from the continuity of $\psi_n (r,\theta)$ in $\theta$, the angular displacement conjugate to $L_z$.  But now let us consider this electron in the reference frame of Fig. 2(b), which rotates at a constant angular speed $\omega$ relative to the initial (nonrotating) reference frame of Fig. 2(a).  Quantum mechanics applies in the rotating reference frame as well; but in the rotating reference frame, as well, the wave function must be single-valued in the angular coordinate.  That is, the transformed wave function $\psi^\prime_n (r, \theta^\prime)$ must be single-valued in $\theta^\prime$.  But then $L_z$ cannot, in general, be a multiple of $\hbar$ in the rotating reference frame!

\begin{figure}
\centerline{
\includegraphics*[width=120mm]{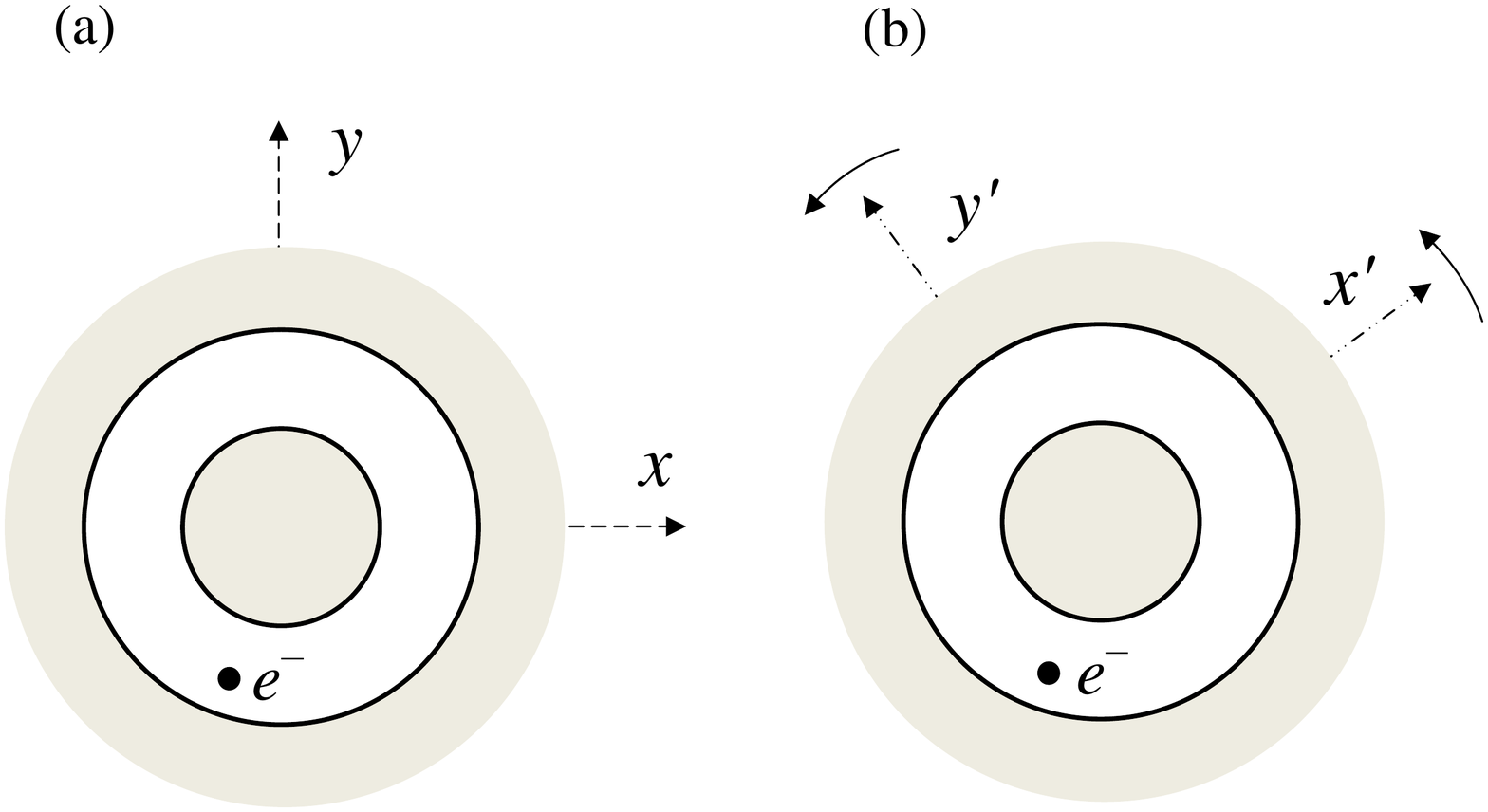}}
\caption[]{(a) The nonrotating reference frame (unprimed axes).  (b) The rotating frame:  the primed axes rotate with angular velocity $\omega$ relative to the unprimed axes.}
\label{Fig2ab}
\end{figure}

We take the Hamiltonian $H= (p_x^2 +p_y^2)/2m + V(x^2 +y^2)$.  Since $e^{iL_z \theta_0/\hbar}$ generates an angular displacement of $\theta_0$, we have $U=e^{iL_z \omega t/\hbar }$ as the unitary transformation from the non-rotating to the rotating reference frame. The transformed state is $\vert \psi^\prime_n\rangle =U\vert \psi_n\rangle$ and the transformed Hamiltonian $H^\prime$ satisfies
\begin{equation}
i\hbar {{d}\over {dt}} \vert \psi^\prime_n \rangle = H^\prime \vert \psi^\prime_n \rangle
\end{equation}
where $H^\prime = UHU^\dagger +i\hbar(dU/dt)U^\dagger$.  For the choice $U=e^{iL_z \omega t/\hbar }$, the transformed Hamiltonian is $H^\prime = H-L_z \omega$; note that $UHU^\dagger=H$, as is easily seen by writing $x^2 +y^2 =r^2$ and
\begin{equation}
p_x^2 +p_y^2= -{\hbar^2 \over r} {\partial \over { \partial r}}\left( r{\partial \over {\partial r}}\right) - {\hbar^2\over {r^2}}{{\partial^2 }\over{\partial \theta^2}}~~~~.
\end{equation}
Thus
\begin{eqnarray}
H^\prime &=& {1\over {2m}} \left( p_x^2 +p_y^2 \right) +V(x^2+y^2)-L_z \omega\cr
&=& {1\over {2m}} \left[ (p_x+m\omega y)^2 +(p_y-m\omega x)^2 \right] - {1\over 2} m\omega^2 (x^2+y^2)+V(x^2+y^2)~~~~.
\end{eqnarray}
We see that quantum mechanics resolves the apparent contradiction by inducing vector and scalar potentials in $H^\prime$.  These potentials are therefore not dispensable.  Yet we distinguish between this example and the others.  In this example, the potentials are an indispensable part of the quantum formalism, in the same way that the Hamiltonian is an indispensable part of the quantum formalism \cite{ramond} (and the potentials are an indispensable part of the Hamiltonian).  But note that locality is not an issue in this (non-inertial) example.

We began by considering an electron in a field without a source. By contrast, our fourth and fifth examples consider sources without fields.

We return to Vaidman's explanation regarding the two counter-rotating charged cylinders, and imagine that they are enclosed by a superconducting shield, which screens them from the field of the passing electron.  It might be argued that, whenever the screening is successful, the AB effect disappears, because the screening current in the superconductor acquires an AB phase of its own that cancels the phase of the electron.  But this argument cannot be correct, as we see from the particular instance of a solenoid containing exactly half of the flux quantum $hc/e$:  in this instance, the Cooper pairs making up the screening current acquire phases that are multiples of 2$\pi$---equivalent to no phase at all---which can never cancel the AB phase of the electron.  Thus the AB effect persists in the absence of any magnetic fields.

It is true that the electric fields of the electron and the screening current remain.  However, as our fifth example we consider the limit of vanishing electric fields.  Let us restrict the electron to an annulus, the inner wall of which is a superconductor that perfectly screens the electron's electric field via an induced current.  This setup corresponds to Fig. 1(a) if the inner wall of the  annulus is the superconductor.  There is no force on the electron due to the magnetic flux $\Phi_B$ threading the annulus, and the electromagnetic field of the electron does not reach the solenoid.  The only force in this example is the force between the electron and the image current it induces in the superconductor.  This force is proportional to $e^2$, while the flux quantum equals $hc/e$; hence, increasing the magnetic flux $\Phi_B$ by $1/e$ makes the AB phase $e\Phi_B /\hbar c$ independent of $e$.  Now suppose we could make the absolute value $e$ arbitrarily small.  In this limit, all forces vanish, while the AB effect is unaffected.  It is true that this scaling of charge and flux is unphysical, in the sense that the value of $e$ is a given; but it is small in reality, and there seems to be no reason not to consider a more general physical setting in which the physical constant $e$ scales along with dynamical variables.  Then we conclude that even if the force between the electron and the screening current vanished asymptotically, the AB effect would persist.

What if we can disconnect the AB phase from the AB effect, such that (in some sense) one appears without the other?  Our final example departs from the topology of the circle and assumes an electron moving at constant velocity in a straight line, passing by a solenoid made of counter-rotating, oppositely charged coaxial cylinders.  According to Ref. \cite{lev}, the electron's magnetic field induces torques on the cylinders that, integrated with respect to the angular displacements of the cylinders, induce a phase in the electron's wave function.  According to quantum mechanics, the phase arises from the vector potential of the solenoid, and implies no change in the velocity of the electron.  By contrast, an explanation via local forces has no place for a vector potential and so the change in the phase of the electron's wave function implies a change in the electron's velocity---a change not seen in experiment.  Here the phase changes without a corresponding physical effect.

In conclusion, we cannot interpret the AB effect as a local effect and at the same time dispense with gauge-dependent quantities. We have elsewhere \cite{ar} considered the AB and related effects as interactions among gauge-invariant quantities (i.e. quantum fields and particles); and only gauge-invariant quantities are measurable.  But then gauge-invariant quantities must interact at a distance:  an electromagnetic field $here$ must affect an electron $there$, etc.  Thus the attempt to dispense with scalar and vector potentials is incompatible with the attempt to interpret the AB effect as a local effect.  Moreover, two of our examples above have no formal description without potentials.  Thus the electromagnetic potentials are, in general, indispensable. In a separate work \cite{acr}, we further elaborate on the inherently nonlocal features of quantum mechanics. We show how any topological effect comprises two components: a continuous effect that fits Vaidman's approach, and a sudden effect that has no local explanation.

\begin{acknowledgments}

Y.A. and E.C. thank the Israel Science Foundation (grant no. 1311/14) for support, and Y.A. acknowledges support also from the ICORE Excellence Center ``Circle of Light", and the German-Israeli Project Cooperation (DIP).  D.R. thanks the John Templeton Foundation (Project ID 43297) and the Israel Science Foundation (grant no. 1190/13) for support.  The opinions expressed in this publication are the authors' and do not necessarily reflect the views of the supporting foundations.

\end{acknowledgments}
\goodbreak

\end{document}